**Effect of cell heterogeneity on isogenic populations with the synthetic genetic toggle switch network: bifurcation analysis of two-dimensional Cell Population Balance Models.**

**Panagiotis Chrysinas, Michail E. Kavousanakis [*], Andreas G. Boudouvis**

National Technical University of Athens, School of Chemical Engineering, Athens 15780, Greece

[*] Corresponding author: mihkavus@chemeng.ntua.gr

**Abstract**

The dynamics of gene regulatory networks are often modeled with the assumption of cellular homogeneity. However, this assumption contradicts the plethora of experimental results in a variety of systems, which designates that cell populations are heterogeneous systems in the sense that properties such as size, shape, and DNA/RNA content are unevenly distributed amongst their individuals. In order to address the implications of heterogeneity, we utilize the so-called cell population balance (CPB) models. Here, we solve numerically multivariable CPB models to study the effect of heterogeneity on populations carrying the toggle switch network, which features nonlinear behavior at the single-cell level. In order to answer whether this nonlinear behavior is inherited to the heterogeneous population level, we perform bifurcation analysis on the steady-state solutions of the CPB model. We show that bistability is present at the population level with the pertinent bistability region shrinking when the impact of heterogeneity is enhanced.

**1. Introduction**

Advances that occurred in the middle of the last and at the beginning of the current century in the fields of biotechnology, genomics and computational biology have supplied us with powerful techniques and methods that can shed light on the complex mechanisms that take place at the single-cell level. However, it is of equally great importance to illuminate and gain profound understanding of the impact that perplex processes, which take place amongst the cells of an isogenic population, have on the individuals' phenotypic variability, and consequently on the average population phenotype. The aforementioned phenomenon -the so-called cell population heterogeneity- has been frequently observed in numerous biological systems. Indicatively, we report the burst variation of bacteriophages (Delbruck, 1945), the existence of transcriptional states of heterogeneity in sporulating cultures of *Bacilus subtillis* (Chung & Stephanopoulos, 1995), various isogenic *Escherichia coli* systems (Elowitz et al., 2002), the endothelial cell surface markers (Oh et al., 2004), transcriptional states at single-cell-resolution (Tischler & Surani, 2013) and single-cell metabolomics (Rubakhin et al., 2013). Furthermore, recent studies have shown the importance of heterogeneity on drug discovery and optimal design of therapeutic strategies (A. Gough et al., 2017; A. H. Gough et al., 2014).

Despite the wealth of experimental evidence for the importance of cell heterogeneity, a number of modeling approaches (e.g., (Chung & Stephanopoulos, 1995; Fedoroff & Fontana, 2002; Sadeghpour et al., 2017) are based on the assumption that all individuals of an isogenic populations share the same phenotype (homogeneous populations). Despite the fact, that this assumption leads to simple mathematical models consisting of systems of ordinary differential equations, disregarding cell heterogeneity can lead to false quantitative predictions, as shown in relative studies (Aviziotis, Kavousanakis, Bitsanis, et al., 2015; Aviziotis, Kavousanakis, & Boudouvis, 2015; Kavousanakis et al., 2009; Mantzaris, 2005; McAdams & Arkin, 1998) in which cell heterogeneity is taken into account.

In this work, we adopt the modeling approach of cell population balance (CPB) to model the dynamics and compute the steady-state solution of heterogeneous isogenic populations, i.e., populations consisting of individuals that carry the same genetic network. In such populations, cell heterogeneity originates from two main sources. The first source, the so-called *intrinsic* heterogeneity, is the result of stochastic fluctuations of regulatory molecules (Alberts et al., 1994), which exist in small concentrations and control a network of intracellular reactions. Gene regulatory molecules are a set of DNA segments inside the cell that interact with each other through their RNA and protein expression products as well as with other intracellular substances. The type and the number of genes expressed at each moment alongside with the intracellular reactions define the phenotype of each cell. Furthermore, gene expression is a stochastic process as shown in (Blake et al., 2003; Elowitz et al., 2002) leading also to phenotypic variability, which originates from intracellular processes

The second source of heterogeneity is the so-called *extrinsic* heterogeneity, which is the result of the uneven distribution of the intracellular content -with the exception of DNA- from a mother cell to its daughter cells during cellular division. The uneven distribution of mother content to the offsprings results in different phenotypes as a result of the different rates of the intracellular reaction network. Furthermore, it is not only the intracellular content which is distributed unevenly; the regulatory molecules are also unevenly distributed, and the phenomenon repeats itself due to the process of cell cycle leading to further phenotypic variability. It has been shown by experimental studies (Elowitz et al., 2002), for *E.coli* populations, that *extrinsic* heterogeneity has a more significant quantitative impact; in this work we focus on the *extrinsic* heterogeneity impact on *E.coli* populations carrying the genetic toggle switch.

In order to quantify the heterogeneity and combine it with the pertinent genetic network, we introduce the CPB models, which were developed in mid 1960s (Eakman et al., 1966; Fredrickson et al., 1967; Tsuchiya et al., 1966). They are partial integro-differential equations and are characterized from high mathematical complexity -even with the application of model-reduction techniques (Stamatakis, 2013). Analytical solutions cannot be obtained for the general case and the use of numerical methods is mandatory, as presented in numerous studies (Liou et al., 1997; Mantzaris et al., 2001a, 2001b, 2001c; Zhang et al., 2003; Zhang et al., 2002; Zhu et al., 2000). However, a common feature of the applied numerical methods is the assumption that the physiological state space boundaries (e.g., the boundaries of the average intracellular content) are known *a priori*. This assumption may be valid for the minimum intracellular content -which we can assume that has the same value from the initial distribution or that is equal to zero- but this does not apply when it comes to the value of the maximum intracellular content. In order to bypass this impediment one can apply a free boundary formulation as presented in (Kavousanakis et al., 2009), based on a valid assumption that the maximum intracellular content is a positive multiple of its average value.

The mathematical formulation of the applied free boundary CPB model is described in Section 2. In particular, we present a two-variable CPB model in order to describe the dynamics of *E.coli* cells carrying a synthetic toggle switch which has been presented in (Gardner et al., 2000). A brief description of its design and mathematical formulation is provided in Section 3. A key feature of this synthetic network is its nonlinear behavior and the existence of a range of extracellular inducer concentration values - IPTG (isopropyl-β-D-thiogalactopyranoside)- with multiple co-existing steady-state phenotypes. In order to examine whether this nonlinear behavior is inherited also to the population level, we first study homogeneous populations using systems of ODEs which describe their dynamics, and the *pseudo* arc-length continuation algorithm (Keller, 1977) to track the entire steady-state solution space as a function of the [IPTG].

The study of heterogeneous populations is presented in Section 4, where we utilize the *pseudo* arc-length method in combination with CPBs, in order to determine and quantify the impact of heterogeneity on the range of bistability (the interval of [IPTG] values with multiple solutions). We need to stress at this point, that the steady-state solution of multivariable CPBs is not a trivial numerical task, with significantly large computational and memory requirements. In order to bypass these difficulties we resort to Newton-like algorithms, and in particular Broyden's algorithm, (Broyden, 1965), which requires only an approximation of the Jacobian matrix, and not the Jacobian matrix itself (as required in Newton-Raphson), thus saving significant computational effort as compared to Newton's method.

In Section 5, we present temporal and steady-state computations for the aforementioned CPB model, which is discretized with the finite element method. We also present the steady-state solution space of heterogeneous populations carrying the synthetic toggle switch as a function of the IPTG concentration. The pertinent bifurcation diagrams show that bistability is also present for heterogeneous cell populations, however the range is narrowed down as the impact of heterogeneity is enhanced. Furthermore, we also study the impact of other parameters on the range of bistability, including the parameters which quantify the asymmetry and sharpness of the division mechanism. Finally, in Section 6 we provide a brief summary of the main results of this study.

## 2. Cell population balance modeling

In this work, we study a two-dimensional CPB model, which describes the dynamics of a heterogeneous population carrying the synthetic toggle switch (Gardner et al., 2000). Each individual of the evolved distribution is characterized by the values of two intracellular variables, namely $x$ and $y$. In the more general case of a $k$-variable CPB model, each cell is characterized by a vector of $k$ intracellular content values, $\underline{x} \equiv (x_1, \ldots, x_k)$, the dynamics of the population are described by the following expression (Mantzaris, 2006):

$$\frac{\partial u(\underline{x},t)}{\partial t} + \nabla_{\underline{x}} \cdot [\underline{R}(\underline{x})u(\underline{x},t)] + \Gamma(\underline{x})u(\underline{x},t) = 2\int_{\underline{x}}^{\underline{x}_{max}} \Gamma(\underline{x}')P(\underline{x},\underline{x}')u(\underline{x}',t)\mathrm{d}^k\underline{x}' - u(\underline{x},t)\int_\Lambda \Gamma(\underline{x})u(\underline{x},t)\mathrm{d}^k\underline{x}, \qquad (2.1)$$

where:

$$\Lambda = [0, x_{1,max}] \times \ldots \times [0, x_{k,max}] \subseteq \mathbb{R}^k, k \in \mathbb{N}, \qquad (2.2)$$

and $\underline{x}_{max} \equiv (x_{1,max}, \ldots, x_{k,max})$ denotes the vector with the maximum intracellular content values. The number density function, $u(\underline{x}, t)$, (Fredrickson et al., 1967), expresses the number of cells with content $\underline{x}$ at time $t$ divided by the total number of cells at this time. The boundary conditions imposed to (2.1) require that the population cells do not grow outside the domain, $\Lambda$, i.e.:

$$u(\underline{0}, t) = u(\underline{x}_{max}, t) = 0. \qquad (2.3)$$

The first term in (2.1) quantifies accumulation, and the second denotes the rate at which cells with intracellular content $\underline{x}$ change their content due to intracellular reactions, $\underline{R}(\underline{x})$. The third term represents division, which yields cells with lower content, when the cell division rate is $\Gamma(\underline{x})$. The first term at the right hand side describes the birth of cells with content, $\underline{x}$, by cells with larger intracellular content. The factor 2 multiplies the integral to model the birth of two cells at the end of each division. The function, $P(\underline{x}, \underline{x}')$, models the mechanism of intracellular content distribution amongst the two daughter cells; in effect $P(\underline{x}, \underline{x}')$ models the probability that a mother cell with content, $x'$, produces a daughter cell with content, $x$, and one of content, $x'-x$. Finally, the last term of the right hand side (dilution term) is the one forcing the solution to reach a steady-state; at this state, the non-normalized distribution of cells reaches a time-invariant shape, while cells continue to proliferate.

Taking into account that $u(\underline{x}, t)$ is the number density function (already normalized by the total number of cells), the following condition must apply:

$$\int_\Lambda u(\underline{x},t)\mathrm{d}^k\underline{x} = 1. \qquad (2.4)$$

Equation (2.1) incorporates single-cell operations through three key functions: $\Gamma(\underline{x})$, $\underline{R}(\underline{x})$ and $P(\underline{x}, \underline{x}')$ known in the relative literature as Intrinsic Physiological State Functions (IPSF). In particular, $\Gamma(\underline{x})$ formulates the single-cell division rate, $\underline{R}(\underline{x})$ describes the single-cell network of reactions containing the net production rates of all intracellular species, and $P(\underline{x}, \underline{x}')$ is the partition probability density function which describes the mechanism of intracellular content partition during the division of a mother cell giving birth to two daughter cells. In this work, we adopt for the division rate, $\Gamma(\underline{x})$, a generalization of the normalized power law, as shown in (Dien, 1994):

$$\Gamma(\underline{x}) = \prod_{i=1}^{k} \left(\frac{x_i}{\langle x_i \rangle}\right)^{m_i}. \tag{2.5}$$

The values of exponents, $m_i$, quantify the sharpness/rapidness of cellular division, with larger values leading to higher division rates.

For the partition probability density function, we assume the simplest possible partition mechanism:

$$P(\underline{x}, \underline{x}') = \sum_{i=1}^{k} \delta(fx_i' - x_i) + \frac{1}{2(1-f)} \delta((1-f)x_i' - x_i), \tag{2.6}$$

where $\delta$ is the Dirac function, and the parameter $f$, quantifies the asymmetry during the cellular division. In every division cycle, the mother cell -with intracellular content $\underline{x}'$- is divided in two daughter cells, with their intracellular content being fractions of $\underline{x}'$, i.e., $f\underline{x}'$ and $(1-f)\underline{x}'$, respectively. The values of parameter $f$ are defined in the interval [0, 0.5], with $f$=0.5 corresponding to the symmetric partition mechanism (equal distribution of mother intracellular content). A more asymmetric mechanism is modeled with lower values of $f$.

2.1 Free boundary transformation

CPB models are characterized by high complexity and in the general case one cannot obtain analytical solutions. A number of numerical methods has been proposed in the relative literature (Liou et al., 1997; Mantzaris et al., 2001a, 2001b, 2001c; Ramkrishna, 2000; Subramanian & Ramkrishna, 1971; Zhang et al., 2003; Zhang et al., 2002; Zhu et al., 2000), however in all these studies the boundaries of intracellular state space are considered *a priori* known. This is a very limiting modeling approach, which can increase the computational demands, since larger domains are required in order to capture accurately the maximum values of the intracellular state space, or they can lead to inaccurate results when the assumption of the size of the domain is invalid. In order to confront this issue, we adopt a free boundary algorithm based on a previous work (Kavousanakis et al., 2009), in which the maximum intracellular content is considered to be a positive multiple of the average intracellular content. When the intrinsic physiological state of each cell content is two-dimensional, i.e., $\underline{x} \equiv (x, y)$, then we consider:

$$\begin{aligned} x_{max} &= \lambda_1 \langle x \rangle \\ y_{max} &= \lambda_2 \langle y \rangle \end{aligned} \tag{2.7}$$

where $\langle x \rangle, \langle y \rangle$ are the average values of intracellular contents $x, y$, respectively. The parameters $\lambda_1, \lambda_2$ are constant positive values. Using Eqs. (2.7) we consider the following transformations for $x, y$, variables:

$$0 \leq x \leq x_{max} \to 0 \leq \frac{x}{\langle x \rangle} \leq \frac{x_{max}}{\langle x \rangle} \xrightarrow{(2.7)} 0 \leq \xi \equiv \frac{x}{\langle x \rangle \lambda_1} \leq 1$$
$$0 \leq y \leq y_{max} \to 0 \leq \frac{y}{\langle y \rangle} \leq \frac{y_{max}}{\langle y \rangle} \xrightarrow{(2.7)} 0 \leq \psi \equiv \frac{y}{\langle y \rangle \lambda_2} \leq 1 \quad (2.8)$$

Using the transformations (2.8), the domain of intrinsic physiological state $\Lambda = [0, x_{max}] \times [0, y_{max}]$ with *a priori* unknown maximum boundaries is transformed to the fixed square domain, $\bar{\Lambda} = [0,1] \times [0,1]$.

The number density function $u(x, y, t)$ is related with the transformed $g(\xi, \psi, t)$ through the relation:

$$u(x, y, t) dx dy = g(\xi, \psi, t) d\xi d\psi \xrightarrow{(2.8)} \lambda_1 \lambda_2 \langle x \rangle \langle y \rangle u(x, y, t) = g(\xi, \psi, t), \quad (2.9)$$

The division rate (2.5) for the two-dimensional case is transformed using (2.8):

$$\Gamma(x, y) = \left(\frac{x}{\langle x \rangle}\right)^{m_1} \left(\frac{y}{\langle y \rangle}\right)^{m_2} \to \Gamma(x, y) = \lambda_1^{m_1} \lambda_2^{m_2} \xi^{m_1} \psi^{m_2} = \lambda_1^{m_1} \lambda_2^{m_2} \tilde{\gamma}(\xi, \psi). \quad (2.10)$$

In addition, we transform the reaction rate vector, $\underline{R}(x, y) = (R_1(x, y), R_2(x, y))$, onto $\underline{\tilde{R}}(\xi, \psi) = (\tilde{R}_1(\xi, \psi), \tilde{R}_2(\xi, \psi))$ and get the transformed (2.1) for two-dimensional intrinsic physiological state, using (2.6)-(2.10):

$$\frac{\partial g}{\partial t} - \frac{1}{\langle x \rangle} \frac{d\langle x \rangle}{dt} \frac{\partial}{\partial \xi}(\xi g) - \frac{1}{\langle y \rangle} \frac{d\langle y \rangle}{dt} \frac{\partial}{\partial \psi}(\psi g) + \frac{1}{\langle x \rangle \lambda_1} \frac{\partial}{\partial \xi}(\tilde{R}_1 g) + \frac{1}{\langle y \rangle \lambda_2} \frac{\partial}{\partial \psi}(\tilde{R}_2 g) +$$
$$\lambda_1^{m_1} \lambda_2^{m_2} g \int_0^1 \int_0^1 \tilde{\gamma} g d\xi d\psi =$$
$$\lambda_1^{m_1} \lambda_2^{m_2} \tilde{\gamma} \left[ \frac{1}{f^{2+m_1+m_2}} g\left(\frac{\xi}{f}, \frac{\psi}{f}\right) + \frac{1}{(1-f)^{2+m_1+m_2}} g\left(\frac{\xi}{1-f}, \frac{\psi}{1-f}\right) - g \right]. \quad (2.11)$$

We need to mention at this point that both average intracellular contents, which appear in (2.11), are unknown values and need to be determined. Taking the first-order moment of (2.1) for each intracellular content variable, and considering the conservation of mass for each of them during cell division the following expressions are derived:

$$\frac{d\langle x \rangle}{dt} = \int_0^{y_{max}} \int_0^{x_{max}} R_1 u dx dy - \langle x \rangle \int_0^{y_{max}} \int_0^{x_{max}} \Gamma u dx dy$$
$$\frac{d\langle y \rangle}{dt} = \int_0^{y_{max}} \int_0^{x_{max}} R_2 u dx dy - \langle y \rangle \int_0^{y_{max}} \int_0^{x_{max}} \Gamma u dx dy \quad (2.12)$$

and using (2.7)-(2.10)

we obtain:

$$\frac{d\langle x \rangle}{dt} = \int_0^1 \int_0^1 \tilde{R}_1 g d\xi d\psi - \lambda_1^{m_1} \lambda_2^{m_2} \langle x \rangle \int_0^1 \int_0^1 \tilde{\gamma} g d\xi d\psi$$
$$\frac{d\langle y \rangle}{dt} = \int_0^1 \int_0^1 \tilde{R}_2 g d\xi d\psi - \lambda_1^{m_1} \lambda_2^{m_2} \langle y \rangle \int_0^1 \int_0^1 \tilde{\gamma} g d\xi d\psi \quad (2.13)$$

$R_1, R_2$ are the single-cell reaction rates for the net production of intracellular contents $x$ and $y$, respectively. Their expressions for the studied synthetic toggle switch are described in the following section.

# 3. Synthetic toggle switch genetic network

A well-known genetic network that has been experimentally designed and implemented in *E. coli* populations is the so-called toggle switch model (Gardner et al., 2000). We choose to study this particular network, motivated by its simplicity and the nonlinear behavior it exhibits at the single-cell level. The genetic toggle is a synthetic bistable network, where flipping between co-existing stable steady-states is feasible using chemical or thermal induction. In the following subsections we provide a brief description of its design, and the respective mathematical framework, for which we apply bifurcation analysis to show the steady-state solution multiplicity of phenotypes within a range of extracellular inducer concentration values (chemical induction).

3.1. Design, principles and mathematical modeling

The toggle switch model consists of two constitutive promoters -genes- and two repressors. The basic mechanism of this model is that each promoter is inhibited by the respective repressor, which has been transcribed by the opposing promoter (i.e., promoter 1 is inhibited by repressor 1, which in turn has been transcribed by promoter 2; promoter 2 is inhibited by repressor 2, which has been transcribed by promoter 1). The inducer induces the respective repressor (i.e., inducer 1 induces repressor 1, and inducer 2 induces repressor 2). The schematic representation of this network is shown in Fig. 1. Schematic of the synthetic toggle switch model(Gardner et al., 2000). (Gardner et al., 2000).

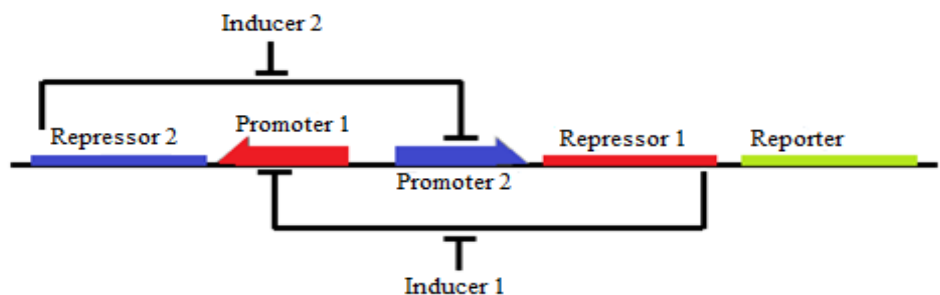

**Fig. 1.** Schematic of the synthetic toggle switch model(Gardner et al., 2000).

The key feature of this network is that it enables the existence of multiple co-existing phenotypes, for the same extracellular conditions, i.e., the same concentration of extracellular inducer. This effect is known as bistability, and in biologic systems is frequently observed, in particular in systems involving the interaction between two genes.

Bistability is of great importance due to the impact that has on basic cell operations, such as the processes of decision making during cell cycle, cell differentiation and apoptosis (Eissing et al., 2004). In addition, bistability is involved with the loss of cellular homeostasis, which in turn is linked to the first stages of cancer (Kim et al., 2007) and the diseases of protein prion (Kellershohn & Laurent, 2001). Moreover, in a -relatively- recent study (Veening et al., 2008) many bistability phenomena such as different phenotypes in clone populations are discussed, which are important for the birth of new species. Bistability has also been observed in virus phage $\lambda$, which contaminates the *E.coli* and can be found in two distinct states: lysogenic and lytic (Tian & Burrage, 2004).

For the case of the synthetic toggle switch, (Gardner et al., 2000) developed a simple mathematical model, which can capture the nonlinear behavior of the toggle switch. In particular, the dynamic behavior of the toggle switch as well as the conditions for bistability can be studied using the following dimensionless model:

$$\frac{dR_1(x,y)}{dt} \equiv \frac{dx}{dt} = \frac{a_1}{1+y^\beta} - \delta x, \tag{3.1}$$

$$\frac{dR_2(x,y)}{dt} \equiv \frac{dy}{dt} = \frac{a_2}{1+\left(\frac{x}{r}\right)^\gamma} - \delta y, \tag{3.2}$$

where $x, y$ are the intracellular concentrations of repressors 1 and 2, respectively, and $r = \left(1 + \frac{[IPTG]}{K}\right)^n$. $[IPTG]$ is the concentration of the extracellular induced IPTG (isopropyl-β-D-thiogalactopyranoside) ([M]), and $K$ is the dissociation constant of IPTG from its bound repressor (*LacR*), and $n$ is the cooperativity of IPTG binding; $a_1$, and $a_2$ are the effective synthesis rates of repressors 1 and 2, respectively; $\beta$ and $\gamma$ are repression cooperativities of promoters 1 and 2, respectively and $\delta$ is the dimensionless degradation rate of repressors 1 and 2.

The first terms of Eqs (3.1)-(3.2) represent the cooperative repression of the constitutively transcribed promoters and the second terms represent the degradation/dilution of the repressors. In this work, we adopt the following set of parameter values as used in the original work of (Gardner et al., 2000): $a_1$=156.25, $a_2$=15.6, $\beta$=2.4, $\gamma$=1 $n$=2.00015, $\delta$=0.005, $K$=2.9618×10$^{-5}$ M, for which the toggle network becomes bistable. Before we proceed with the examination of the effect of cell heterogeneity, we first present the parametric analysis results for the homogeneous isogenic population with cells carrying the genetic toggle.

3.2. Parametric analysis of homogeneous cell population

In this case, all cells of the population are identical in the sense that they all feature the same phenotype, i.e., their intracellular content is the same, and equal with the average intracellular content. Thus, the number density function of a homogeneous population can be represented using the Dirac function (Kavousanakis et al., 2009):

$$u(x,y) = \delta(x - \langle x \rangle, y - \langle y \rangle). \tag{3.3}$$

Substituting (3.3) in (2.12) yields the following equations for the average intracellular content, $\langle x \rangle$ and $\langle y \rangle$:

$$\frac{d\langle x \rangle}{dt} = R_1(\langle x \rangle, \langle y \rangle) - \langle x \rangle \Gamma(\langle x \rangle, \langle y \rangle) \xrightarrow{(2.5),(3.1)} \frac{d\langle x \rangle}{dt} = \frac{a_1}{1+\langle y \rangle^\beta} - (\delta + 1)\langle x \rangle, \tag{3.4}$$

$$\frac{d\langle y\rangle}{dt} = R_2(\langle x\rangle, \langle y\rangle) - \langle y\rangle \Gamma(\langle x\rangle, \langle y\rangle) \xrightarrow{(2.5),(3.2)} \frac{d\langle y\rangle}{dt} = \frac{a_2}{1+\left(\frac{\langle x\rangle}{r}\right)^\gamma} - (\delta+1)\langle y\rangle. \qquad (3.5)$$

The complete solution space of steady-state phenotypes of homogeneous isogenic populations carrying the toggle switch (i.e., $\frac{d\langle x\rangle}{dt} = \frac{d\langle y\rangle}{dt} = 0$) is computed for a range of [IPTG] values using the so-called *pseudo* arc-length continuation algorithm (Keller, 1977). In Fig. 2. Dependence of steady-state average intracellular concentration of (a) repressor 1, ⟨x⟩, and (b) repressor 2, ⟨y⟩, for a homogeneous population carrying the toggle switch on the IPTG concentration. By increasing the repression cooperativity of promoter 2, β, the range of bistability is enlarged. Lines with circles, rectangles and triangles correspond to β=2.2, 2.3 and 2.4, respectively. Solid and dashed lines depict stable and unstable steady-state solutions, respectively. we present the results of this analysis for different values of promoter 1 repression cooperativity, $\beta$. One can observe the existence of a range of IPTG concentration values within which three different steady-state solutions co-exist, with two of them being dynamically stable (upper and lower branches of solutions), and one dynamically unstable (intermediate branch, marked with dashed lines in Fig. 2. Dependence of steady-state average intracellular concentration of (a) repressor 1, ⟨x⟩, and (b) repressor 2, ⟨y⟩, for a homogeneous population carrying the toggle switch on the IPTG concentration. By increasing the repression cooperativity of promoter 2, β, the range of bistability is enlarged. Lines with circles, rectangles and triangles correspond to β=2.2, 2.3 and 2.4, respectively. Solid and dashed lines depict stable and unstable steady-state solutions, respectively.). Furthermore, by increasing the value of parameter, $\beta$, the range of bistability is enlarged.

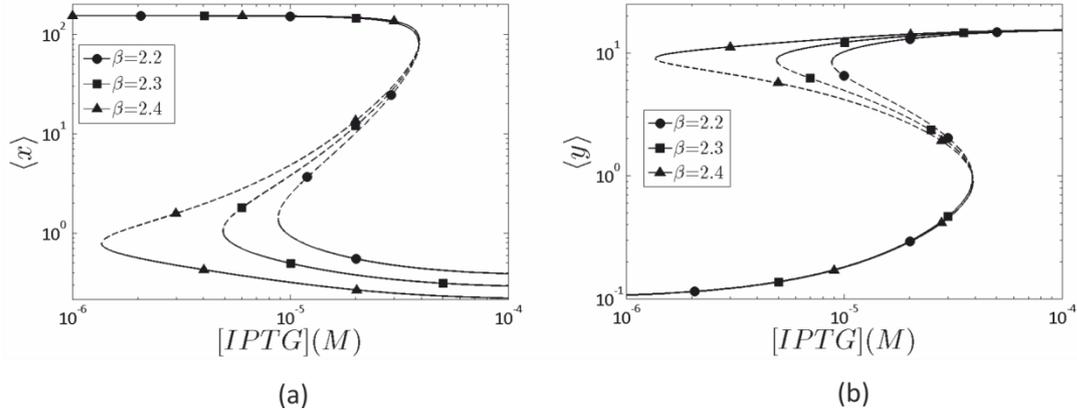

**Fig. 2.** Dependence of steady-state average intracellular concentration of (a) repressor 1, ⟨x⟩, and (b) repressor 2, ⟨y⟩, for a homogeneous population carrying the toggle switch on the IPTG concentration. By increasing the repression cooperativity of promoter 2, β, the range of bistability is enlarged. Lines with circles, rectangles and triangles correspond to β=2.2, 2.3 and 2.4, respectively. Solid and dashed lines depict stable and unstable steady-state solutions, respectively.

Thus, bistability which is observed at the single-cell level (Gardner et al., 2000) is also inherited to the population level when all individuals feature identical phenotypes. However, neglecting heterogeneity can lead to significant quantitative, as well as qualitative discrepancies even for the average intracellular properties (Aviziotis, Kavousanakis, Bitsanis, et al., 2015; Aviziotis, Kavousanakis, & Boudouvis, 2015; Kavousanakis et al., 2009; Mantzaris, 2005; McAdams & Arkin, 1998). In order to study the effect of cell heterogeneity, one needs to perform bifurcation analysis on the CPB model (see (2.1) or the transformed (2.11)). This analysis is not as trivial as for the homogenous population case, in which ordinary differential equations are only involved.

Here, we employ the finite element method for the discretization of the CPB model using the software package COMSOL Multiphysics® (COMSOL, 2017). The finite element discretization yields a nonlinear set of algebraic equations, which we solve with iterative pseudo-Newton algorithms, and in particular Broyden's algorithm (Broyden, 1965), with significantly lower computational and memory requirements compared to the standard Newton-Raphson method. In order to implement Broyden's algorithm, as well as bifurcation analysis techniques, we utilize the interface of COMSOL with MATLAB as presented in (Kavousanakis et al., 2009).

## 4. Bifurcation analysis of heterogeneous cell populations

In general, the discretization of a boundary value problem, using the finite element method yields a system of nonlinear equations, which we denote with:

$$\underline{F}(\underline{u},\rho)=\underline{0} \tag{4.1}$$

where $\rho$ is the bifurcation parameter (e.g., for our study the [IPTG]), and $\underline{u}$ is the discretized form of the sought solution. The nonlinear set of equations (4.1) can be iteratively solved using the Newton-Raphson algorithm, which at each step solves the linearized set of equations:

$$\frac{\partial \underline{F}(\underline{u}^{(n)})}{\partial \underline{u}} \delta \underline{u}^{(n)} = -\underline{F}(\underline{u}^{(n)}), \tag{4.2}$$

where $\underline{J}^{(n)} \equiv \frac{\partial \underline{F}(\underline{u}^{(n)})}{\partial \underline{u}}$, is the Jacobian matrix computed at step, $n$, of the algorithm; the solution of (4.2), $\delta \underline{u}^{(n)}$, is used to update the solution vector for the next iteration, $n+1$, using: $\underline{u}^{(n+1)} = \underline{u}^{(n)} + \delta \underline{u}^{(n)}$.

For CPBs the Jacobian matrix is a highly dense matrix, and the required computational cost to obtain all entries of it, as well as the memory requirements to store them are extremely high. For a typical case of a 10,000-dimensional $\underline{u}^{(n)}$ vector, it would require the computation and storage of $10^8$ entries for the Jacobian matrix or 800 MB for each iteration of the Newton-Rapshon method, since the algorithm does not facilitate the re-use of data from previous iterations.

4.1 Broyden's algorithm.

An alternative is to modify the Newton-Raphson method so that approximate partial derivatives are used for the computation of the Jacobian at an expense of slower convergence. In particular, one can adopt quasi-Newton methods, which maintain approximations of the solution $\underline{u}^*$ and the Jacobian at the solution $\frac{\partial F(\underline{u}^*)}{\partial \underline{u}}$ as the iteration progresses. If we denote with $\underline{u}_c$ and $\underline{J}_c$ the current approximate solution and Jacobian, then the solution is updated according to the relation:

$$\underline{u}_+ = \underline{u}_c - \underline{J}_c^{-1} \underline{F}(\underline{u}_c). ) \tag{4.3}$$

After the computation of $\underline{u}_+$, $\underline{J}_c$ is also updated to form the approximate Jacobian for the next step, $\underline{J}_+$; the relation used to construct $\underline{J}_+$ determines the quasi-Newton method. In this work, we adopt Broyden's algorithm, which is locally superlinearly convergent, hence is a very powerful alternative of Newton-Rapshon method, and computes $\underline{J}_+$ using:

$$\underline{J}_+ = \underline{J}_c + \frac{\underline{F}(\underline{u}_c)(\underline{u}_+ - \underline{u}_c)^T}{(\underline{u}_+ - \underline{u}_c)^T(\underline{u}_+ - \underline{u}_c)}. \tag{4.4}$$

Broyden's algorithm is an example of a secant update, which means that the approximation for $\underline{J}_+$ satisfies the secant equation: $\underline{J}_+(\underline{u}_+ - \underline{u}_c) = \underline{F}(\underline{u}_+) - \underline{F}(\underline{u}_c)$. However, for problems with a dense Jacobian matrix Broyden's algorithm (4.4) is still ineffective. In order to sidestep the computational and memory limitations imposed by the density of the Jacobian matrix, we utilize the Sherman-Morrison approach, which minimizes the storage requirements and incorporates only a small number of vectors for the update of the Jacobian's approximations. In particular, the algorithm reads (Kelley, 1995):

---

1. $r_0 = \left\| \underline{P}^{-1}\underline{F}(\underline{u}_0) \right\|_2, n = 0$

   $\underline{s}_0 = -\underline{P}^{-1}\underline{F}(\underline{u}_0), itc = 0.$

2. Do while $itc < maxit$:

   (a) $n = n + 1; itc = itc + 1$

   (b) $\underline{u}_n = \underline{u}_n + \underline{s}_{n-1}$

   (c) Evaluate $\underline{P}^{-1}\underline{F}(\underline{u}_n)$

   (d) if $\left\| \underline{P}^{-1}\underline{F}(\underline{u}_n) \right\|_2 \leq \tau_r r_0 + \tau_a$ exit.

   (e) if $n < nmax$ then

        i. $\underline{z} = -\underline{P}^{-1}\underline{F}(\underline{u}_n)$

        ii. for $j = 0, n-1$

   $$\underline{z} = \underline{z} + \underline{s}_{j+1}\underline{s}_j^T\underline{z}/\left\| \underline{s}_j \right\|_2^2$$

        iii. $\underline{s}_{n+1} = \dfrac{\underline{z}}{1 - \dfrac{\underline{s}_n^T \underline{z}}{\left\| \underline{s}_j \right\|_2^2}}$

   (f) if $n = nmax$ then

   $n = 0, \underline{s}_0 = -\underline{P}^{-1}\underline{F}(\underline{u}_n)$

---

This variant of Broyden's algorithm uses restarting, and the storage is cleared when a maximum number, $nmax$, of stored vectors, $\underline{s}_j$ is reached; $maxit$ is the maximum number of Broyden's iterations, and $\tau_r, \tau_a$ are the relative and absolute tolerances which determine the termination criterion of the iterative process. The matrix $\underline{P}$ is the preconditioner applied in order to achieve better convergence rate. In this work, we use as preconditioner the Jacobian of the CPB equation at $\underline{u}_0$, and removing the entries resulting from the contribution of the integral term, $\lambda_1^{m_1}\lambda_2^{m_2} g \int_0^1 \int_0^1 \tilde{\gamma} g d\xi d\psi$, in Eq. (2.11). Thus, we only compute a sparse banded matrix, which can be used as preconditioner. We report here, that according to Broyden's algorithm reported above, the only expensive operation is the solution of the sparse linear system, $\underline{z} = -\underline{P}^{-1}\underline{F}(\underline{u}_n)$, which is performed only once per iteration.

Broyden's algorithm can be trivially extended for the application of the *pseudo* arc-length continuation method (Keller, 1977). In particular, we introduce one additional unknown variable, the bifurcation parameter, $\rho$ (here $\rho$ is the $IPTG$ concentration value). Thus, the sought solution vector is the augmented, $\underline{u}_{augm} = [\underline{u}\,\rho]^T$. The residual vector, $\underline{F}$ is augmented by one additional constraint, $N$, the so-called arc-length constraint. In this work, we adopt the following formulation for constraint, $N$:

$$N = \left(\frac{\underline{u}_{\rho_1}-\underline{u}_{\rho_0}}{S_1-S_0}\right)^T \left(\underline{u}-\underline{u}_{\rho_1}\right) + \frac{\rho_1-\rho_0}{S_1-S_0}(\rho-\rho_1) - dS, \qquad (4.5)$$

where $\underline{u}_{\rho_0}, \underline{u}_{\rho_1}$ are computed solutions at parameter values $\rho_0$ and $\rho_1$, respectively; $S_1 - S_0$ is the Euclidean distance between the augmented vector solutions $[\underline{u}_{\rho_1}\,\rho_1]$ and $[\underline{u}_{\rho_0}\,\rho_0]$, and $dS$ is the (user-selected) arc-length parameter step for the performance of the continuation algorithm. The results of the Broyden's algorithm implementation in conjunction with the *pseudo* arc-length continuation method are presented in the following section.

## 5. Results and discussion

In this section, we present the results of the bifurcation analysis implemented on the CPB model (2.11) in order to study the effect of heterogeneity on the range of bistability for cell populations carrying the synthetic toggle switch network. The computation of the steady-state solution of a nonlinear problem requires the application of iterative algorithms, and their success can be guaranteed when a good initial estimation of the sought solution is provided. This initial estimation can be obtained through transient simulations.

5.1 Transient simulations

The two-dimensional CPB model (2.11) is discretized (as reported above) using the finite element method. Among the different time integration techniques we choose the application of an explicit time-integrator, and in particular the 4[th] order Runge-Kutta type, whose basic advantage is the low requirements of computational memory. In Fig. 3. Snapshots of the number density function, $u(x,y)$, computed from the simulation of (2.11)-(2.13), in combination with (3.1)-(3.2) when the initial distribution is a shifted bivariate Gaussian distribution (5.1). , we present snapshots of a transient simulation for the following set of parameter values: $[IPTG] = 16.5 \times 10^{-5}M, m_1 = m_2 = 2, f = 0.5, \lambda_1 = \lambda_2 = 5$. The computational domain, $\bar{\Lambda} = [0,1] \times [0,1]$, is discretized using 4088 triangular elements, and the resulting number of degrees of freedom using quadratic basis functions is 8297.

The initial condition for the transient simulation presented in Fig. 3. Snapshots of the number density function, $u(x,y)$, computed from the simulation of (2.11)-(2.13), in combination with (3.1)-(3.2) when the initial distribution is a shifted bivariate Gaussian distribution (5.1). is a shifted bivariate Gaussian distribution:

$$u(x,y,t=0) = \frac{1}{2\pi\sigma_1\sigma_2}\exp\left(-\left(\frac{(x-\mu_1)^2}{2\sigma_1^2}+\frac{(y-\mu_2)^2}{2\sigma_2^2}\right)\right), \qquad (5.1)$$

where $\mu_1, \mu_2$ are the mean values of intracellular contents $x$ and $y$, respective at $t=0$, and $\sigma_1, \sigma_2$ the standard deviation of $x$ and $y$. The initial distribution of the normalized density function, $g$, is the following:

$$g(\xi, \psi, t = 0) = \frac{\lambda_1 \lambda_2 \mu_1 \mu_2}{2\pi \sigma_1 \sigma_2} \exp\left(-\left(\frac{(\lambda_1 \mu_1 \xi - \mu_1)^2}{2\sigma_1^2} + \frac{(\lambda_2 \mu_2 \psi - \mu_2)^2}{2\sigma_2^2}\right)\right). \tag{5.2}$$

The snapshots in Fig. 3. Snapshots of the number density function, $u(x, y)$, computed from the simulation of (2.11)-(2.13), in combination with (3.1)-(3.2) when the initial distribution is a shifted bivariate Gaussian distribution (5.1). show the evolution of the number density function, $u(x, y)$, which is obtained from the normalized $g(\xi, \psi)$ through (2.8)-(2.9). The transient simulation shown in Fig. 3. Snapshots of the number density function, $u(x, y)$, computed from the simulation of (2.11)-(2.13), in combination with (3.1)-(3.2) when the initial distribution is a shifted bivariate Gaussian distribution (5.1). also shows that the initial shifted Gaussian distribution evolves towards a wider steady-state solution with shorter amplitude at dimensionless time, $t=6$, at which practically a steady-state solution has been reached. Similar transient simulations have been performed for different IPTG concentration values, and $f$ parameter values (which quantify the degree of asymmetry during division) for large time intervals, at the end of which the evolved distributions converge to steady-state solutions.

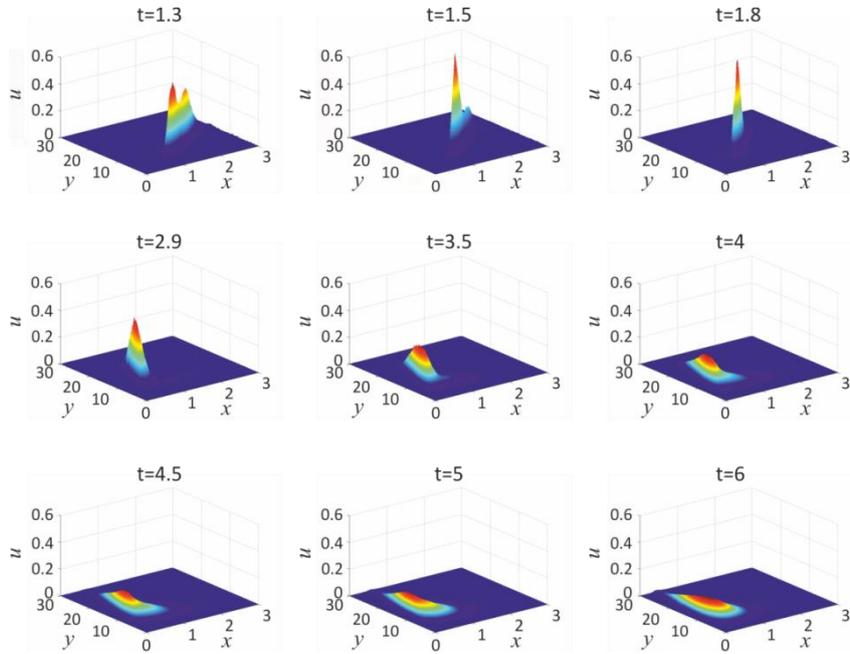

**Fig. 3.** Snapshots of the number density function, $u(x, y)$, computed from the simulation of (2.11)-(2.13), in combination with (3.1)-(3.2) when the initial distribution is a shifted bivariate Gaussian distribution (5.1). Parameter set values: $[IPTG] = 16.5 \times 10^{-5} M$, $m_1 = m_2 = 2$, $f = 0.5$, $\lambda_1 = \lambda_2 = 5$, $\mu_1 = \mu_2 = 5$, and $\sigma_1 = \sigma_2 = 0.25$.

In practice, a steady-state solution can be reached when transient simulations are performed for time intervals, $t \in [0,100]$. In Fig. 4. Steady-state solutions of the bivariate distribution $u(x, y)$ as obtained from long transient simulations, $t \in [0,100]$. Parameter set values: $m_1 = m_2 = 2$, $\lambda_1 = \lambda_2 = 5$. The initial distribution is a bivariate Gaussian distribution (5.1), with, $\mu_1 = \mu_2 = 5$, and $\sigma_1 = \sigma_2 = 0.25$., we present the converged steady-state number density function distributions after long transient simulations for different [IPTG] and $f$ values.

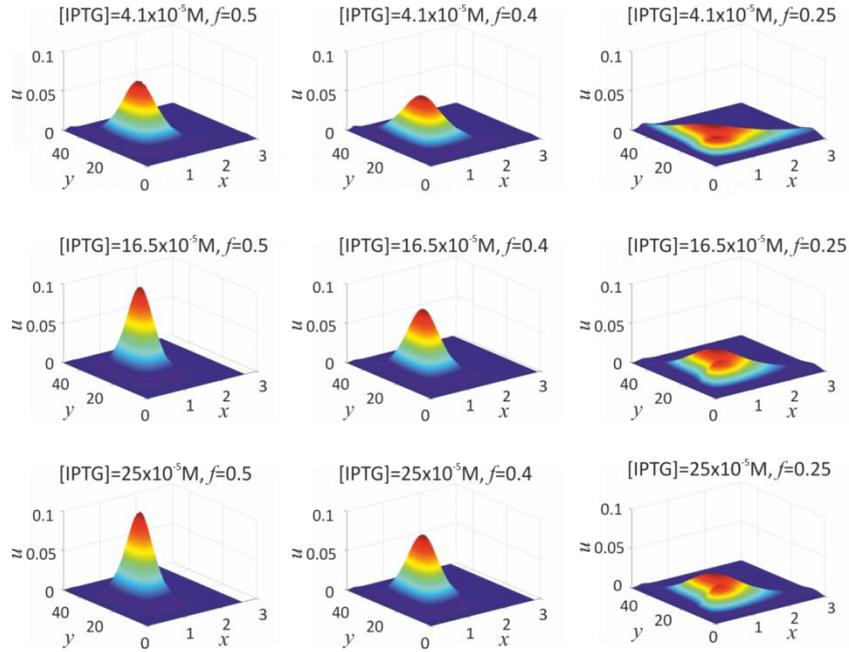

**Fig. 4.** Steady-state solutions of the bivariate distribution $u(x,y)$ as obtained from long transient simulations, $t \in [0,100]$. Parameter set values: $m_1 = m_2 = 2$, $\lambda_1 = \lambda_2 = 5$. The initial distribution is a bivariate Gaussian distribution (5.1), with, $\mu_1 = \mu_2 = 5$, and $\sigma_1 = \sigma_2 = 0.25$.

By increasing the extracellular inducer concentration value, [IPTG], the steady-state shape, $u(x,y)$ becomes wider (spreads over larger intervals of intracellular content). By decreasing the value of parameter, $f$, i.e., increasing the degree of heterogeneity during cell division, one can observe major quantitative, as well as, qualitative differences in the steady-state solution shape. In particular, by decreasing the value of parameter, $f$ the deviation of the steady-state distribution becomes wider and interestingly enough, for the lower presented values of $f$=0.25, the obtained distribution becomes two-humped.

We re-iterate, that the computed steady-state solutions presented in Fig. 4. Steady-state solutions of the bivariate distribution $u(x,y)$ as obtained from long transient simulations, $t \in [0,100]$. Parameter set values: $m_1 = m_2 = 2$, $\lambda_1 = \lambda_2 = 5$. The initial distribution is a bivariate Gaussian distribution (5.1), with, $\mu_1 = \mu_2 = 5$, and $\sigma_1 = \sigma_2 = 0.25$. are obtained through the performance of long transient simulations. However, utilizing this approach to compute the entire solution space for different parameter values is inadequate especially for systems that exhibit bistability; a large number of different initial conditions should be tested for different sets of parameter values in order to obtain the different steady-state solutions, which can co-exist within -*a priori* unknown- intervals of parameter values. A more systematic way should be adopted and in particular we utilize bifurcation analysis techniques, e.g. the *pseudo* arc-length parametric continuation algorithm (Keller, 1977), which enables the computation of all possible co-existing steady-state solutions, both dynamically stable and unstable. The *pseudo* arc-length continuation method is implemented in conjunction with Broyden's algorithm (Broyden, 1965) to override computational and memory limitations.

5.2 Bifurcation analysis results

Initially, we present results of this analysis for different values of parameter $f$ (asymmetry factor of mother content distribution during division). In Fig. 5. Dependence of the steady-state average intracellular , we present a typical bifurcation diagram illustrating the dependence of the average intracellular content, $\langle x \rangle$, on the concentration values of the extracellular inducer. The solid lines correspond to stable steady-state solutions, whereas dashed lines show unstable steady-state solutions. In order to have a clear picture of the impact of heterogeneity on the bistability range, we also include in Fig. 5. Dependence of the steady-state average intracellular the results obtained from the homogenous population case. One can observe that bistability is also present for heterogeneous populations, however its range is significantly reduced when the impact of heterogeneity is enhanced (by lowering, $f$, values).

Starting from low IPTG concentration values, i.e., $[IPTG] < 1 \mu M$, there is only one stable steady-state distribution, featuring high $\langle x \rangle$, and low $\langle y \rangle$ concentration values. By gradually increasing $[IPTG]$, a critical turning point is encountered at approximately $[IPTG] = 40 \mu M$, for all cases of studied heterogeneous and homogenous populations (with small discrepancies between them), which initiates a transition towards steady-state distributions with low intracellular content, $\langle x \rangle$ (and respectively high $\langle y \rangle$). In a reverse experiment, where the initial IPTG concentration is high, i.e., $[IPTG] > 100 \mu M$ one can observe only states with low content, $\langle x \rangle$. By gradually reducing $[IPTG]$, a transition towards states with high $\langle x \rangle$ concentration values, will be initiated sooner when the degree of heterogeneity is high. By decreasing the value of parameter $f$, the range of bistability shrinks, and in particular for $f = 0.25$ the bistability interval spans over a very narrow interval of IPTG values. As the degree of heterogeneity is further reduced, the range of bistability is enlarged and reaches its maximum for homogenous populations.

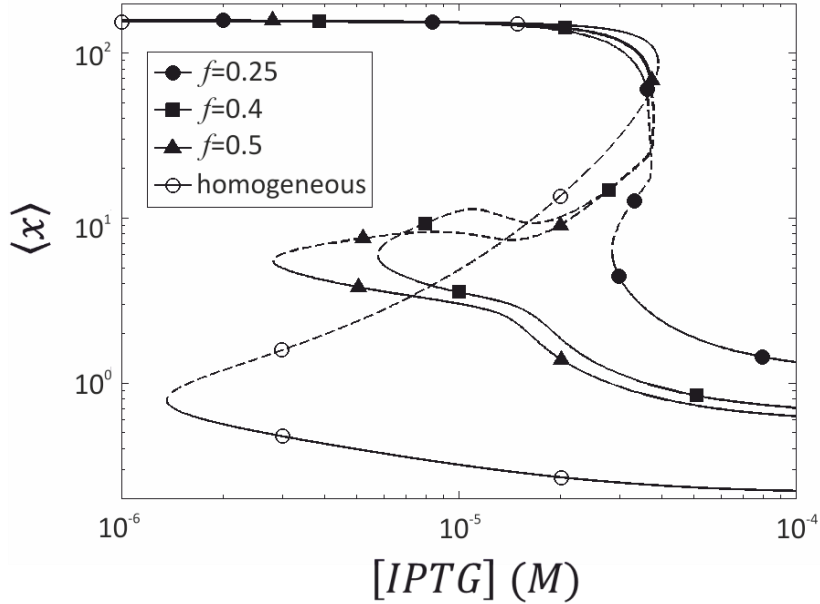

**Fig. 5.** Dependence of the steady-state average intracellular concentration, $\langle x \rangle$, on the IPTG concentration values. The lines with solid circles, rectangles and triangles correspond to heterogeneous populations with $f = 0.25, 0.4$ and $0.5$, respectively. The line with open circles corresponds to homogeneous cell populations. Solid and dashed lines represent stable and unstable steady-state solutions, respectively. Parameter set values: $m_1 = m_2 = 0.35$, $\lambda_1 = \lambda_2 = 5$.

In order to quantify the stability of the obtained steady-state solutions, one can compute the spectrum of eigenvalues of the Jacobian matrix; however as reported above the computation and storage of Jacobian matrix which is derived by discretized multidimensional CPB problems is a prohibitive task, since the Jacobian in such cases is a highly dense matrix. Alternatively, one can study the stability of the obtained steady-state solutions by performing temporal simulations with slightly perturbed extracellular conditions (IPTG concentration values). In particular, we use as initial condition the steady-state solution computed at a parameter value [IPTG]; then we perturb the value of [IPTG] and initiate the simulation to observe the evolution of the solution. If the solution converges to a nearby one, then it is characterized as dynamically stable; on the other hand if it diverges to a solution which is far from the initial distribution then it is characterized as dynamically unstable. In Fig. 6. Time evolution of the average intracellular contents $\langle x \rangle, \langle y \rangle$ starting from three steady-state (upper, lower, and intermediate branch) solutions of $u(x, y)$ computed at $[IPTG] = 10^{-5}M$. The extracellular inducer is slightly increased by $10^{-6}M$. The line with filled, and open circles shows the evolution of $\langle x \rangle$ and $\langle y \rangle$, respectively. Small perturbations to the (a) upper and (b) lower solution branches lead to nearby solutions, whereas the slightly perturbed (c) intermediate solution converges to the upper solution branch signifying the instability of the intermediate branch solution. we present the stability analysis performed on three co-existing steady-state solutions obtained for $[IPTG] = 10^{-5}M$ and the following set of parameter values: $f = 0.5, m_1 = m_2 = 0.35, \lambda_1 = \lambda_2 = 5$. The extracellular condition is slightly perturbed by setting $[IPTG] = (10^{-5} + 10^{-6})M$ and we initiate a long transient simulation. Starting from the solutions which belong to the upper and lower branches, we observe that the average intracellular content converge finally to a nearby solution, and the same also holds for the cell density function. On the other hand, when the initial condition is the solution belonging to the intermediate branch, we observe the solution diverging from its initial state and finally converging to the upper solution branch. This behavior is typical for dynamically unstable solutions.

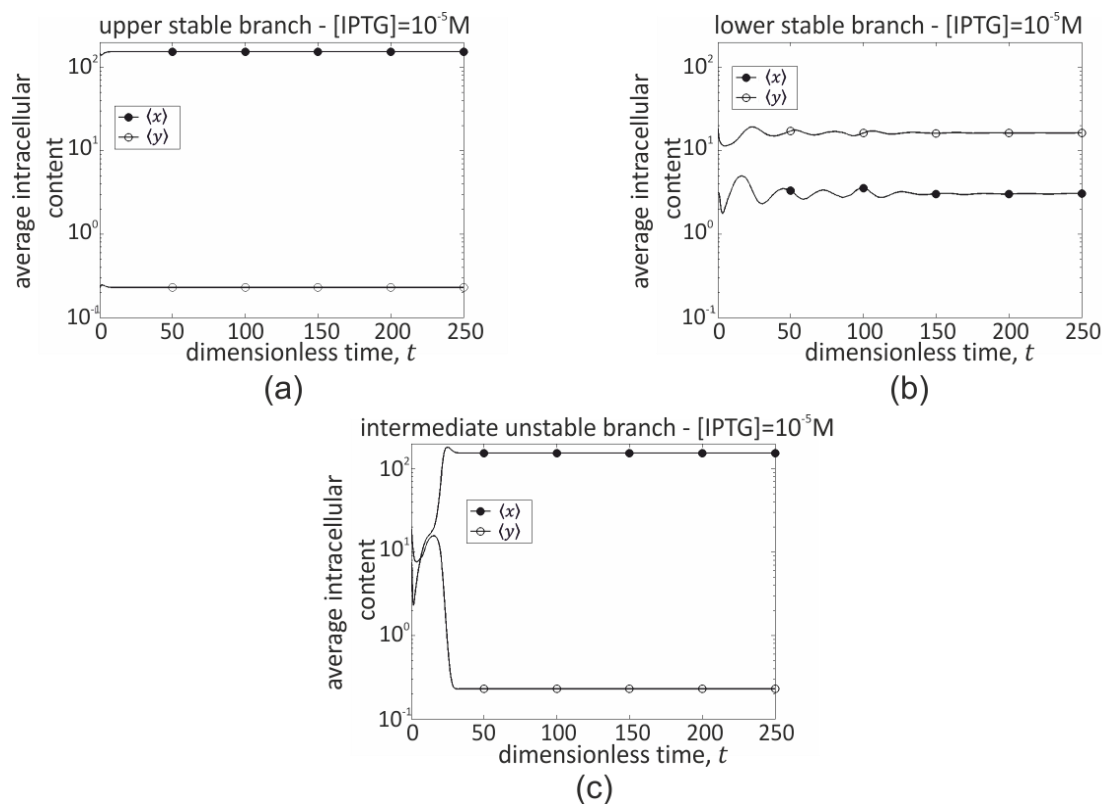

**Fig. 6.** Time evolution of the average intracellular contents $\langle x \rangle, \langle y \rangle$ starting from three steady-state (upper, lower, and intermediate branch) solutions of $u(x,y)$ computed at $[IPTG] = 10^{-5}M$. The extracellular inducer is slightly increased by $10^{-6}M$. The line with filled, and open circles shows the evolution of $\langle x \rangle$ and $\langle y \rangle$, respectively. Small perturbations to the (a) upper and (b) lower solution branches lead to nearby solutions, whereas the slightly perturbed (c) intermediate solution converges to the upper solution branch signifying the instability of the intermediate branch solution.

In addition to the previous bifurcation analysis for different values of the parameter $f$, we also perform the same analysis to study the effect of different parameter values that are incorporated in the CPB model. In Fig. 7, we show the effect of the division rate sharpness, which is quantified through parameters $m_1$, and $m_2$ (division rate becomes larger by increasing the values of exponents $m_1$, and $m_2$). One can observe that the change in division rate has little effect on the position of the right turning point value, which signifies transitions from populations featuring high $\langle x \rangle$ average concentration values, towards populations with low average $\langle x \rangle$ phenotypes. However, it has significant impact on the bistability range, when an inverse transition is attempted; in particular, by increasing the sharpness rate the bistability range is enlarged, and the position of the left turning point is located at lower $[IPTG]$ values.

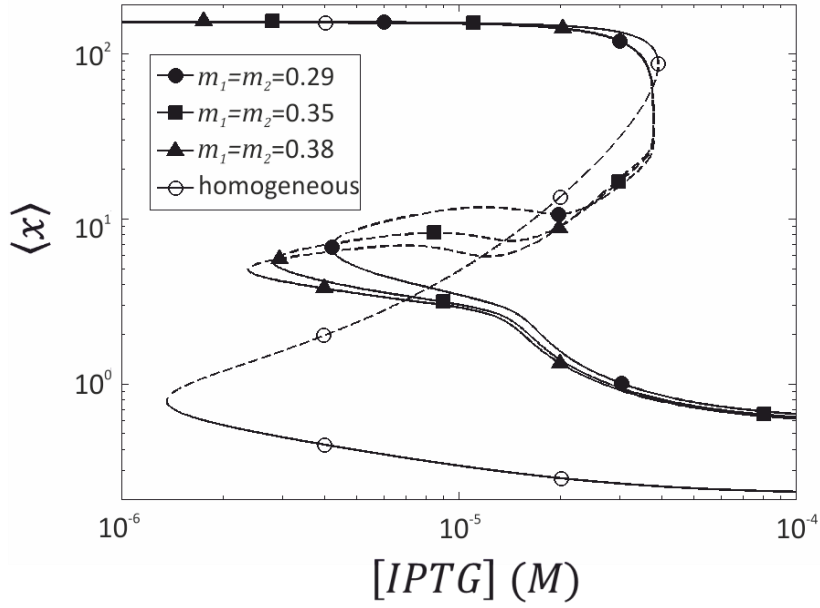

**Fig. 7.** Dependence of the steady-state average intracellular content, $\langle x \rangle$ on the extracellular inducer concentration, $[IPTG]$ for different single-cell division rates. Lines with solid circles correspond to lower division rate ($m_1 = m_2 = 0.29$); lines with solid rectangles correspond to intermediate division rate ($m_1 = m_2 = 0.35$) and lines with solid triangles depict solutions for higher division rate ($m_1 = m_2 = 0.38$). Lines with open circles corresponds to homogeneous populations. Parameter values for the solution of CPB: $f = 0.5$, $\lambda_1 = \lambda_2 = 5$.

Finally, we examine the effect of promoter 1 repression cooperativity, $\beta$, on the range of bistability range, as illustrated in Fig. 8. . The position of the right turning point remains practically invariant for all examined cases of parameter value, $\beta$. On the other hand, the left turning point which signifies transitions from states of low average $\langle x \rangle$ phenotype, towards states of $\langle x \rangle$ values, moves towards lower $[IPTG]$ concentration values as $\beta$ value increases. We report that the same behavior is also observed for homogeneous populations (see Fig. 2), however in all cases of $\beta$ values the range of bistability shrinks in comparison with the ones observed for homogeneous populations.

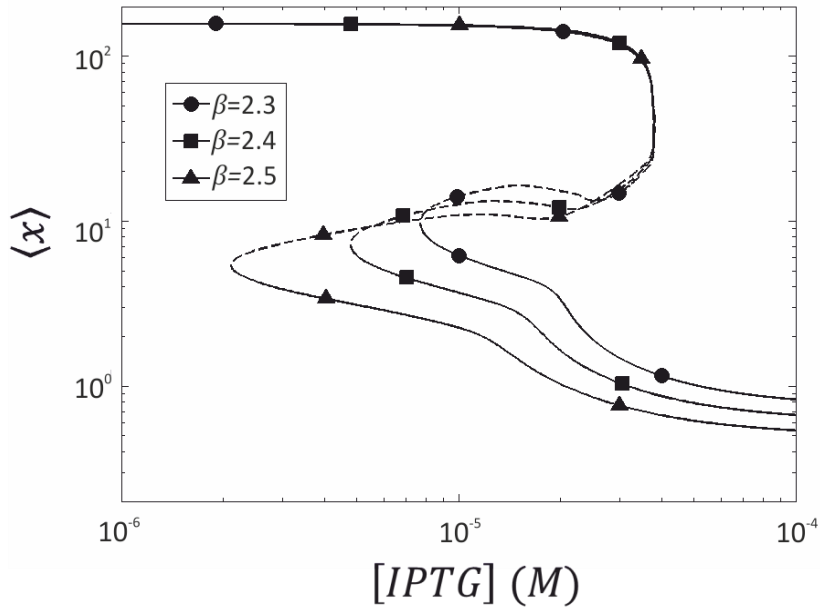

**Fig. 8.** Dependence of the steady-state average intracellular content, $\langle x \rangle$ on the extracellular inducer concentration, $[IPTG]$ for values of the promoter 2 repression cooperativity, $\beta$. Lines with solid circles correspond to $\beta = 2.3$; lines with solid rectangles correspond to $\beta = 2.4$ and lines with solid triangles depict solutions for higher $\beta$ values ($\beta = 2.5$). Parameter values for the solution of CPB: $f = 0.5$, $\lambda_1 = \lambda_2 = 5$.

**Conclusions**

    The primary goal of this work is to study the impact of heterogeneity on the steady-state (time invariant) phenotype of isogenic populations featuring nonlinear behavior. In particular, we study isogenic populations carrying the synthetic toggle switch, which at the single-cell exhibits phenotypic multiplicity within a certain range of extracellular conditions, e.g., extracellular inducer concentration.

    Cell heterogeneity is incorporated by adopting the CPB modeling approach, which consists of partial-integro-differential equations. These models can describe the effect of the so-called extrinsic heterogeneity, which is present due to the uneven distribution of cell content amongst the daughter cells during division. In addition, the intracellular reaction network is also incorporated and the division rate of each individual of the population. Among the different numerical methods existing for the numerical solution of CPBs, we adopt a finite element method framework, combined with a free boundary formulation as shown in (Kavousanakis et al., 2009), which surpasses numerical impediments originating from the fact that the physiological state space boundaries are not known *a priori*. This formulation is based on the assumption that the maximum intracellular content is a multiple of the average content, which is computed on the fly along with the solution of the CPB equation.

    We demonstrate this modeling approach on cell populations which carry the synthetic toggle switch, where each individual is characterized by two intracellular variables (two-dimensional CPB equation). Since we are interested in studying the steady-state behavior of such populations as a function of the extracellular environment, we need to utilize iterative algorithms for the solution of large sets of nonlinear equations, which are derived from the discretization of CPBs using the finite element method. However, the standard Newton-Raphson iteration algorithm is

practically non-applicable for the case of multidimensional CPBs, since it requires at each iteration the computation, storage, and treatment of the Jacobian matrix, which in this case is a highly dense matrix.

An alternative approach is to solve nonlinear sets of equations with pseudo-Newton algorithms, which use only approximations of the Jacobian at the cost of convergence rate. The method used in this work is the Broyden's algorithm (Broyden, 1965), which updates the Jacobian matrix at each iteration using its estimation at the previous step. We can further reduce the computational cost by applying the Sherman-Morrison approach (Kelley, 1995) which utilizes and stores only a small number of vectors, in order to update of the approximate Jacobian matrix, and the sought solution vector.

Broyden's algorithm is combined with the *pseudo* arc-length parametric continuation algorithm (Keller, 1977) in order to track the entire solution space, including both stable and unstable solutions. By adopting this numerical framework we are able to produce bifurcation diagrams, which clearly depict a range of extracellular conditions within which multiple steady-state solutions can co-exist, as also shown for the simplified case of homogeneous populations. The comparison of heterogeneous and homogeneous populations, shows clearly that as we enhance the effect of heterogeneity the bistability interval shrinks, and the transition between states of utterly different phenotypes is more rapid. In particular, when we study the dependence of phenotypes on the extracellular inducer IPTG concentration we observe that the low end of bistability interval tends towards higher IPTG values as we increase the effect of heterogeneity. By performing the same analysis for different sets of parameter values, we also examine the effect of cell division rate, which shows that higher division rates yield wider bistability range.

Thus, it is evident that cell heterogeneity is a critical factor that needs to be addressed in all modeling approaches for the simulation of biological systems. Neglecting its effect can lead to false quantitative and qualitative predictions. In this paper, we emphasize on the study of the extrinsic heterogeneity which can be addressed with the use of CPBs. It is also of high interest to examine the effect of the second type of heterogeneity, the so-called intrinsic heterogeneity, which originates from stochastic noise during intracellular reactions. This study stands beyond the scope of the current work, and requires the development of stochastic models as presented in (Aviziotis, Kavousanakis, Bitsanis, et al., 2015; Aviziotis, Kavousanakis, & Boudouvis, 2015). In particular, kinetic Monte Carlo algorithms can be developed in order to simulate the effect of intrinsic parameters that cannot be incorporated through deterministic modeling.